\documentclass[runningheads]{llncs}
\usepackage{graphicx}
\usepackage{authblk}
\usepackage{hyperref}
\usepackage{booktabs} 
\usepackage{multirow}
\usepackage{amssymb}
\usepackage{amsmath}
\usepackage[T1]{fontenc}
\usepackage{lmodern}
\usepackage{graphicx}
\usepackage{float}
\usepackage[misc]{ifsym}
\usepackage{subcaption}

\begin{document}

%
\title{Multi-modal Mood Reader: Pre-trained Model Empowers Cross-Subject Emotion Recognition
}
\titlerunning{Multi-modal Mood Reader}
%

%
%
%
\author{Yihang Dong\inst{1,2}\orcidID{0000-1111-2222-3333} \and
Xuhang Chen\inst{2} \and
Yanyan Shen\inst{1,2} \and
Michael Kwok-Po Ng\inst{3}\and
Tao Qian \inst{4} \and
Shuqiang Wang\inst{1,2}\orcidID{0000-0003-1119-320X}\textsuperscript{(\Letter)}}

\authorrunning{Y. Dong et al.}
%
\institute{University of Chinese Academy of Sciences \\ \and
Shenzhen Institutes of Advanced Technology, Chinese Academy of Sciences\\ \and
 Department of Mathematics, Hong Kong Baptist University\\ \and
Faculty of Innovation Engineering, Macau University of Science and Technology\\
\email{sq.wang@siat.ac.cn}}
\maketitle              

\begin{abstract}
Emotion recognition based on Electroencephalography (EEG) has gained significant attention and diversified development in fields such as neural signal processing and affective computing. However, the unique brain anatomy of individuals leads to non-negligible natural differences in EEG signals across subjects, posing challenges for cross-subject emotion recognition. While recent studies have attempted to address these issues, they still face limitations in practical effectiveness and model framework unity. Current methods often struggle to capture the complex spatial-temporal dynamics of EEG signals and fail to effectively integrate multimodal information, resulting in suboptimal performance and limited generalizability across subjects. To overcome these limitations, we develop a Pre-trained model based Multimodal Mood Reader for cross-subject emotion recognition that utilizes masked brain signal modeling and interlinked spatial-temporal attention mechanism. The model learns universal latent representations of EEG signals through pre-training on large scale dataset,  and  employs Interlinked spatial-temporal attention mechanism to process Differential Entropy(DE) features extracted from EEG data. Subsequently, a multi-level fusion layer is proposed to integrate the discriminative features, maximizing the advantages of features across different dimensions and modalities. Extensive experiments on public datasets demonstrate Mood Reader's superior performance in cross-subject emotion recognition tasks, outperforming state-of-the-art methods. Additionally, the model is dissected from attention perspective, providing qualitative analysis of emotion-related brain areas, offering valuable insights for affective research in neural signal processing.

\end{abstract}
\section{Introduction}
Brain-computer interface (BCI) systems have long been an aspirational goal for researchers in the fields of computer science, neuroscience, and psychology. The envisioned maturation of BCI technology is expected to significantly expand human sensory, cognitive, and operational capabilities, offering unprecedented depth and breadth in human-machine interaction. However, truly efficient human-machine interaction relies not solely on the machine's ability to interpret and execute human commands, but more critically, on the sensitive detection and accurate recognition of users' implicit emotional states. Consequently, the task of emotion recognition has naturally emerged as a key research area.


Although emotion recognition methods based on various physiological signals each have their unique characteristics, they predominantly face challenges related to the complexity of signal collection, difficulties in data processing, and high costs. Therefore, non-invasive EEG, with its relatively low cost, convenient signal collection, superior signal representation capability, and non-harmful nature to subjects, has rapidly become a primary research focus in the field of emotion recognition. The array of electrodes placed on the scalp effectively collects signals reflecting brain electrical activity. Through precise analysis and processing of these signals, an individual's emotional state can be effectively revealed.

Non-invasive EEG signals are not without flaws. The unique natural physiological and anatomical structures of each individual introduce various degrees and aspects of noise interference into the measured EEG, imparting non-stationary characteristics to it. Additionally, issues such as the non-Euclidean distribution of multi-channel EEG electrodes based on biological topography collectively impact the accuracy of cross-subject emotion recognition tasks. Researchers have attempted to tackle these challenges from different directions. Transfer learning, as an indirect approach, has been utilized to migrate emotion recognition models, originally trained and adapted for existing subjects, to new individuals, aiming to minimize the EEG differences between the source and target domains~\cite{li2019multisource,yan2023inspiration}. Although this approach has indeed achieved certain recognition effects, it does not fundamentally solve the problem of cross-subject emotion recognition. Considering the graph-like topological structure of EEG channels and the rapid development of Graph Neural Networks (GNN), a surge of cross-subject emotion recognition methods based on GNN has emerged~\cite{li2022gmss,zhong2020eeg}, attempting to capture the local and global relationships among EEG channels. Similarly, other methods have also achieved certain yet limited improvements in recognition accuracy~\cite{li2022cross,wang2018bone}.

Motivated by the recent emergence of pre-trained models and their outstanding performance in downstream tasks~\cite{chen2023seeing,ortega2023brainlm,Luo_Chen_Chen_Li_Wang_Pun_2024,Li_2023_ICCV,ijcai2023p129}, we recognized that the high-dimensional semantic information of EEG extracted by encoders trained on large-scale subject-independent datasets may contain global generic representations beneficial for emotion recognition tasks. Simultaneously, considering the DE feature, which has been proven to be the most effective individual computational characteristic for EEG-based emotion recognition tasks~\cite{du2020efficient,tao2020eeg,shen2022contrastive}, as well as eye movement features often used as an additional modality to aid emotion recognition tasks and confirmed to improve recognition accuracy~\cite{liu2021comparing}, we propose Mood Reader, a novel multi-modal and cross-scale fusion model for cross-subject emotion recognition that integrates global generic representations of EEG and the spatio-temporal interaction information in specific DE features. Specifically, our contributions are as follows:
\begin{enumerate}
\item We propose an emotion recognition model architecture that integrates multi-modal and cross-scale information, demonstrating exceptional performance in cross-subject recognition tasks. This architecture also proves that encoders pre-trained on large-scale EEG data possess the ability to learn emotion-related features to a certain extent.
\item We have designed an attention-based interlinked spatio-temporal module for learning the compensatory relationships between spatio-temporal information, which aids in the fusion of spatio-temporal features.
\item Supported by extensive experiments, we provide a biologically plausible interpretation of emotion recognition research based on EEG and make reasonable hypotheses.
\end{enumerate}
\section{Related Works}
\subsection{EEG-Based Emotion Recognition}
Research on EEG-based emotion recognition has attracted significant attention from researchers in recent years. EEG signals, due to their inherent attribute of directly measuring the electrical activity on the scalp surface to capture changes in brain neural activity, have almost become the most powerful data type for emotion recognition. With the development of deep learning technologies that automatically extract features from data, there has been increasing focus on their application in critical areas such as computer vision, natural language processing, and emotion recognition. Given that EEG signals are essentially multi-channel time series signals, to reasonably utilize this nature, Li et al. proposed a BILSTM network framework based on multimodal attention, which is used to learn the best temporal characteristics, and input the learned deep features into the DNN to predict the emotional output probability of each channel~\cite{li2020exploring}. Since emotions are the comprehensive result of the human body's response to external stimuli~\cite{ledoux1989cognitive}, researchers naturally began to use attention mechanisms, similarity coordination constraints, and other multimodal fusion methods to integrate single-modal features extracted from EEG, eye movement signals, facial expressions, etc., from different perspectives for emotion recognition research~\cite{jiang2023multimodal,vazquez2022emotion,jia2021hetemotionnet,ma2019emotion,chaparro2018emotion,zheng2018emotionmeter}. The ablation studies of these researches powerfully validate the correctness of the direction of multimodal data fusion in the field of emotion recognition.
\subsection{Masked Brain Signal Modeling}
With the rapid development of self-supervised pre-training models in the fields of computer vision and natural language processing, researchers have migrated this technology, which can learn generic knowledge representations for target tasks, to multiple fields including brain signal decoding. MBSM, proposed by Chen et al.~\cite{chen2023seeing}., is a self-supervised learning model for large-scale fMRI datasets, which helps its encoder learn the general representation in fMRI signals through the learning process of repeatedly reconstructing complete data from unmasked fMRI signals, further adapting the encoder to different downstream tasks using simple fine-tuning techniques~\cite{chen2023seeing,chen2024cinematic}. Meanwhile, Bai et al. successfully overcame the inherent variability and noise of EEG data by deeply mining the semantics of EEG signals over time, migrated this technology to EEG, and applied it to downstream tasks such as decoding high-resolution images from brain activity~\cite{bai2023dreamdiffusion}.
\subsection{Spatial-Temporal Attention Mechanism}
Complex neural activities are often effectively achieved by the synergistic collaborative processing of multiple sets of continuous neural signals across various distinct neural regions.~\cite{yang2023default,rollo2023dynamical,you2022fine,gong2023generative,wang2020ensemble,hu20233,pan2021characterization,wang2009variational}. Therefore, for research fields including brain decoding and neural information processing, it is crucial to simultaneously capture the temporal and spatial information in brain signal data, for instance, fMRI and EEG. Since the attention mechanism was proposed, various application fields of deep learning, including but not limited to the field of emotion recognition based on EEG signals, have made great progress, and many excellent spatiotemporal information extraction modules have been deduced~\cite{cherian2020spatio,ahn2023star,zhou2022transvod}. In the realm of EEG-based emotion recognition, the Spatial-Temporal Attention (STA) mechanism emerges as a notable innovation for enhancing the interpretability and performance of deep learning models.  This type of architecture ingeniously integrates spatial and temporal dimensions of EEG signals through parallel attention pathways, enabling the model to concurrently learn spatial correlations across EEG channels and temporal dependencies within signal sequences. Li et al. employed the spatio-temporal combination network R2G-STNN, which contains local-global feature combing, to extract the intrinsic information of EEG signals~\cite{li2019regional}. In order to mine the spatiotemporal information related to emotional judgment, Gong et al. designed a stacked parallel spatial and temporal attention streams to respectively extract the spatial features and temporal features of the specially processed EEG signals~\cite{gong2023astdf}. Although previous researchers have obtained satisfactory results, most of them have ignored the interaction and mutual compensation between spatial information and temporal information in EEG signals. In these network streamlines, the two kinds of information are often ignored. Parallel offload processing, which is contrary to the processing flow of neural signals in complex neural activities, may not be enough to handle more complex neural signal processing tasks.
\section{Methodology}

\begin{figure}[ht]
\centering
\includegraphics[width=\linewidth]{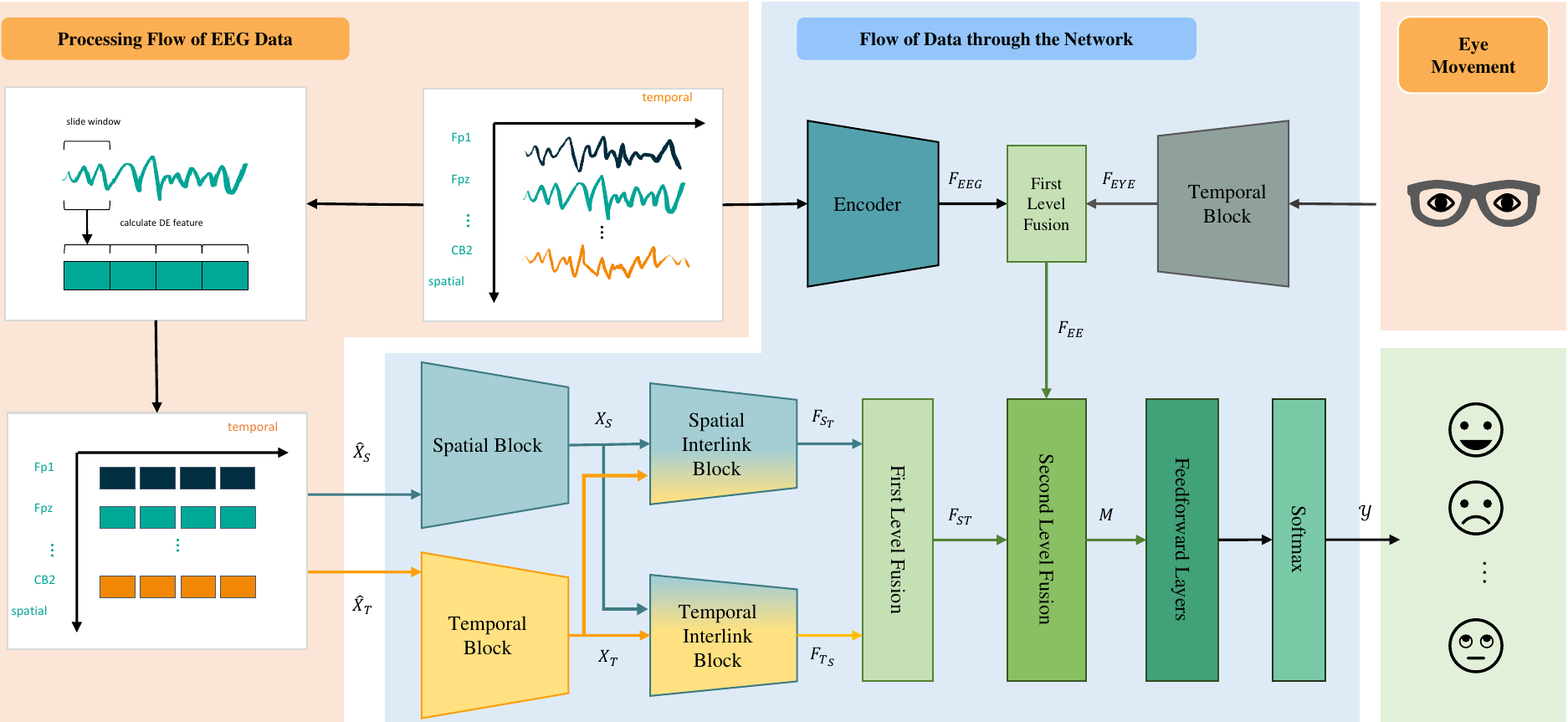}
\captionsetup{belowskip=-30pt}
\caption{The overall architecture of our proposed model and the way related data flows in it.}
\label{fig:overview}
\end{figure}

\subsection{Overview}
Mood Reader is an emotion recognition model that accommodates multi-modal, cross-scale information and successfully integrates these features, its overall architecture is illustrated in Figure \ref{fig:overview}. The model encompasses three distinct types of input. EEG monitoring data, subjected to simple preprocessing, are encoded by an encoder pre-trained on a large-scale dataset, resulting in outputs that contain rich semantic representations. An attention-based interlinked spatial-temporal mechanism captures the intrinsic spatio-temporal information from DE features extracted from EEG data. Additionally, a set of similar temporal attention blocks analyzes corresponding eye movement features, aiming to complement the shortcomings of EEG data. The acquired features are progressively fused in a sequential manner, ensuring that the model genuinely learns the comprehensive complementarity between different modal and scale information, and utilizes it for emotion recognition tasks.
\subsection{MBSM Based EEG General Representation Learning}
Due to the inherent brain differences among subjects and external noise affecting the signals collected by non-invasive EEG, we adopted a pre-training technique known as masked brain signal modeling, which has been proven effective multiple times, to learn meaningful and contextually rich general knowledge representations from cross-subject, noisy, large-scale EEG data~\cite{bai2023dreamdiffusion,chen2023seeing,chen2024cinematic}. Specifically, we completed this task by training an autoencoder-decoder with an asymmetric architecture similar to that in~\cite{bai2023dreamdiffusion}on the EEG Motor Movement/Imagery Dataset~\cite{schalk2004bci2000}. In this model, the temporal signals of EEG data were divided into tokens of a specific size, where a larger ratio of tokens would be randomly masked, and the architecture-simple decoder had to reconstruct the EEG data using the remaining unmasked tokens arrangement, combined with semantically rich embeddings outputted by the encoder after processing the original EEG data.The performance of masked reconstruction improves and reaches a peak when the mask ratio hits 75\% ~\cite{bai2023dreamdiffusion,chen2023seeing,chen2024cinematic}. Consequently, by removing the decoder from this trained model, we obtained an encoder with excellent capability in extracting general EEG representations.
\subsection{Attention Based Interlinked Spatial-Temporal Mechanism}
Deep learning analyses of neural processes typically begin with various types of feature extraction. For complex neural activities like EEG-based emotion recognition, which exhibit high spatiotemporal continuity, it's crucial to accurately unearth intrinsic spatial and temporal features and their interrelations to obtain more valuable information. To address this challenge, this section introduces the interlinked spatial-temporal attention module for processing DE features, comprised of multiple parallel spatiotemporal blocks and interactive spatiotemporal blocks (as depicted in Figure.), these blocks collectively facilitate the extraction of spatio-temporal information and enable the communication and complementation between the extracted spatial and temporal features.
\subsubsection{Spatial and Temporal Representation of DE features}
In the domain of EEG-based emotion recognition, the DE feature, which quantifies the variability of EEG signals, has been proven to be the most effective feature, capturing brain activities related to emotions. It itself has channel-related explicit spatial information and implicit temporal information compressed within a single sliding window.In order to make the spatiotemporal information in the DE feature more balanced, we expand the number of sliding windows in the DE feature to the explicit temporal dimension, obtaining$\hat{X} \in \mathbb{R}^{N \times F \times C}$, where $N$denotes the number of sliding windows involved in the DE feature computation. For further dimension transformation, we get$\hat{X}_S \in \mathbb{R}^{C \times (N \cdot F)}$for spatial information representation and $\hat{X}_T \in \mathbb{R}^{N \times (C \cdot F)}$ for temporal information representation.
\subsubsection{Parallel Spatiotemporal Feature Extraction}
To capture the dynamically varying key information, we apply dedicated spatial and temporal attention blocks to the differentiated DE feature spatial representation$\hat{X}_S$ and temporal representation$\hat{X}_T$, respectively. Specifically, for the spatial representation$\hat{X}_S \in \mathbb{R}^{C \times (N \cdot F)}$, an initial layer normalization is employed to to yield $\hat{X}_S$, which effectively mitigates the internal covariate shift within the spatial representation data, thereby maintaining the stability of its distribution, as follows,
\begin{equation}
X_S^\prime = \text{LayerNorm}(\hat{X}_S)
\end{equation}
After the normalization, multi-head attention (MHA) computation is implemented on $X_S^\prime$. Within head $i$, $X_S^\prime$ is processed by three separate linear networks, transforming the input into different representational spaces to obtain the corresponding query $Q_S^i$, key $K_S^i$, and value $V_S^i$, denoted as,
\begin{equation}
Q_S^i = X_S^\prime W_S^{q_i}, \quad K_S^i = X_S^\prime W_S^{k_i}, \quad V_S^i = X_S^\prime W_S^{v_i}
\end{equation}
Where $W_S^{q_i}$, $W_S^{k_i}$, $W_S^{v_i}$ are the learnable network parameters respectively.

Based on the attention calculation method, the attention output for each head $i$ is obtained as $A_S^i = \text{Attention}(Q_S^i, K_S^i, V_S^i)$. With some processing, the DE feature space representation's MHA output $A_S$ can be obtained as $A_S = \text{Concat}(A_S^1, A_S^2, \ldots, A_S^h)W_S^{\text{MHA}}$, where $h$ is the number of attention heads, and $W_S^{\text{MHA}}$ is a linear mapping weight. The attention calculation is described as,
\begin{equation}
\text{Attention}(Q_S^i, K_S^i, V_S^i) = \text{softmax}\left(\frac{Q_S^i {K_S^i}^T}{\sqrt{d_k}}\right) V_S^i
\end{equation}
where ${d_k}$ represents the dimension of the key.

At the conclusion of the spatial attention block, a residual connection is introduced by adding the dropout-processed MHA output $A_S$ to $X_S^\prime$. This combined result is then subjected to another layer normalization to produce the final spatial representation of the DE feature  $X_S$
 , which is within the space $\mathbb{R}^{C \times (N \cdot F)}$. The calculation process of this part is as follows,
\begin{equation}
X_S = \text{LayerNorm}(\text{Dropout}(A_S) + X_S')
\end{equation}

Similarly, we process $\hat{X}_T \in \mathbb{R}^{N \times (C \cdot F)}$ with a structurally analogous temporal attention block to obtain the temporal representation of the DE feature $X_T \in \mathbb{R}^{N \times (C \cdot F)}$

\begin{figure}[htbp]
\captionsetup{belowskip=-10pt}
\centering
\includegraphics[width=0.5\textwidth]{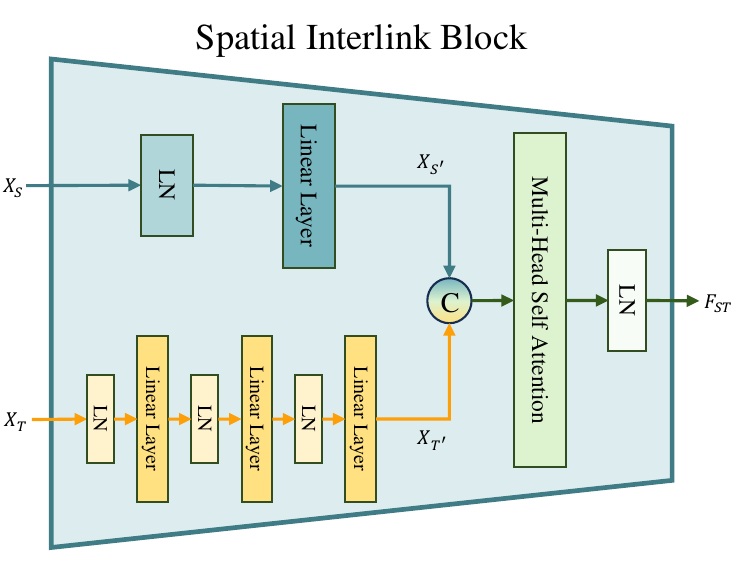}
\caption{Overview of the spatial interlink block, the temporal feature $X_T$ undergos multiple transformations to align with the spatial feature $X_S$, and after concatenation, the interlink process is completed through multi-head attention computation.}
\label{fig:interlink SB}
\end{figure}

\subsubsection{Interlink Between Spatio-Temporal Block} For the spatial features $X_S$ and temporal features $X_T$, which are processed by the spatio-temporal attention block, they are fed into the corresponding interlink block to facilitate the intersection of primary and secondary dimensions. Specifically, the interlinked spatial block will receive both $X_S$ and $X_T$. However, for the primary information $X_S$, which has already achieved a high degree of spatial information integration after being processed by the spatial attention block, it only requires a single linear transformation to become $X_{S^\prime}$, in order to preserve its existing information. As for $X_T$, it undergoes multiple transformations in an attempt to align with the spatial feature dimensions, resulting in $X_{T^\prime}$. Subsequently, we concatenate these two feature parts and perform MHA calculation to obtain the spatial features $F_{S_T}$ that are interlinked with the temporal dimension, as shown in Figure~\ref{fig:interlink SB}. Similarly, we can also obtain the temporal features $F_{T_S}$ that are interlinked with the spatial dimension.
\subsection{Multi-Level Score Filtering for Feature Fusion}
Previous sections presented outputs from various modules, each designed for emotion recognition through different pathways, yielding high-dimensional features. To address potential redundancy and the needs of multimodal cross-scale fusion, we introduce a multi-level fusion layer based on attention mechanisms. This module synergizes and refines features by highlighting relevant information and filtering out redundancy, thus improving multimodal emotion recognition efficacy.

Prior to the commencement of the fusion process, the MBSM latent representation, spatial-temporal representation, and eye movement representation, are projected onto a unified dimensional space through a series of transformations, starting with layer normalization, followed by flattening of the feature vectors, and culminating in a linear transformation. As a result of these operations, a unified feature representation $F \in \mathbb{R}^{D_{\text{unified}}}$ is obtained, where $D_{\text{unified}}$ denotes the dimensionality of the unified feature space.

In the first level of fusion, specifically for the interlinked spatial-temporal features $F_{S_T}$ and $F_{T_S}$, preprocessing is executed utilizing a linear layer and layer normalization, which preserves their dimensional attributes. Subsequently, inspired by~\cite{jiang2023multimodal,zuo2024prior,zuo2021multimodal}, a simplified cross-attention mechanism is employed to delineate the intrinsic spatial-temporal relationships between the two, thereby augmenting the interactivity of the internal information. Specifically, the attention scores $\text{Score}_s$ and $\text{Score}_t$, which represent the spatial features taking into account temporal features and time features taking into account spatial features respectively, are transformed using the softmax function to ascertain the corresponding fusion weights $c_s$ and $c_t$,
\begin{equation}
\text{Score}_s=(F_{S_T}W), \text{Score}_t=(F_{T_S}W)
\end{equation}
\begin{equation}
c_{s}, c_{t} = \text{softmax}(\text{Score}_s, \text{Score}_t),
\end{equation}
where $W_S$,$W_T\in \mathbb{R}^{D_{\text{unified}}×D_{\text{unified}}}$is a learnable weight matrix, the process of obtaining $c_{t}$ is analogous to this. The spatial-temporal fusion features $F_{ST}$ are then calculated by,
\begin{equation}
{F_{ST}} = c_s F_{S_T} + c_t F_{T_S}
\end{equation}
The processing of the MBSM latent representation and eye movement representation, corresponding to $F_{EEG}$ and $F_{EYE}$ respectively, is conducted in a manner akin to the aforementioned methodology. This approach yields an additional fused feature $F_{EE}$, computed as a weighted sum: $F_{EE} = c_{eeg}F_{EEG} + c_{eye}F_{EYE}$, where $c_{eeg}$ and $c_{eye}$ represent the fusion weights derived from the respective attention scores.
In the terminal fusion layer, an integration is requisite for the synthesized features ${F_{ST}}$ and $F_{EE}$, which respectively represent the interlinked spatio-temporal information emanating from the DE feature and the comprehensive information spanning multiple modalities and scales.The features that have undergone layer normalization are concatenated and subsequently processed through self-attention computation, yielding the final integrated feature representation $M$,
\begin{equation}
M = \text{Attention}\left(\text{Concat}\left(F_{ST}, F_{EE}\right)\right)
\end{equation}
For the feature $M$, an initial batch normalization is employed, followed by the deployment of a classifier comprised of three linear layers and a softmax function, which outputs predictions for the emotional labels $y$. The discrepancy between the predicted emotional labels and the true emotional labels $\hat{y}$ is quantified using the cross-entropy loss function,
\begin{equation}
\mathcal{L} = -\frac{1}{N} \sum_{i=1}^{N} \sum_{c=1}^{C} \hat{y}_{ic} \log(y_{ic})
\end{equation}
wherein $N$ represents the batch size, and $C$ designates the count of label categories.

\begin{figure}[htbp]
\centering
\begin{subfigure}[b]{0.18\textwidth} 
    \includegraphics[width=\textwidth]{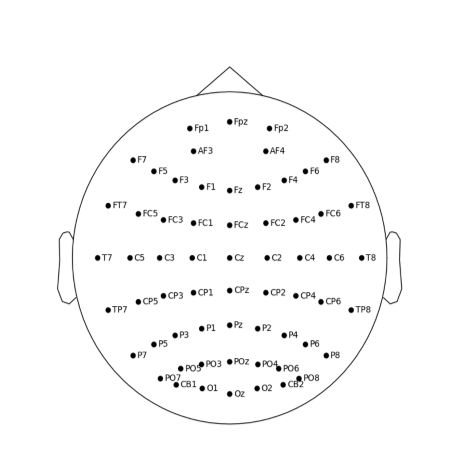}
    \caption{}
    \label{fig:image1}
\end{subfigure}
\hfill 
\begin{subfigure}[b]{0.18\textwidth}
    \includegraphics[width=\textwidth]{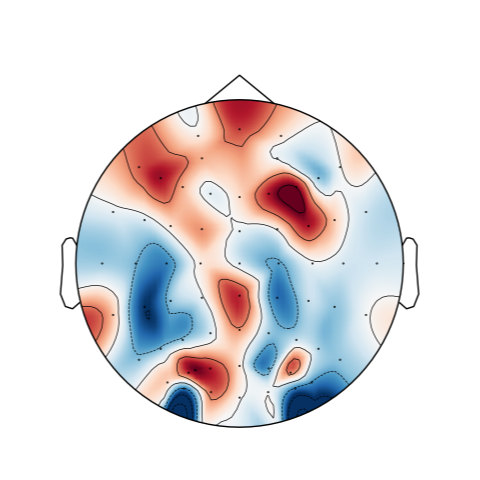}
    \caption{}
    \label{fig:image2}
\end{subfigure}
\hfill
\begin{subfigure}[b]{0.18\textwidth}
    \includegraphics[width=\textwidth]{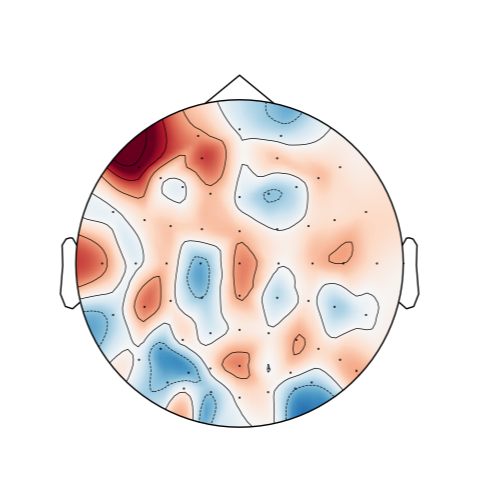}
    \caption{}
    \label{fig:image3}
\end{subfigure}
\hfill
\begin{subfigure}[b]{0.18\textwidth}
    \includegraphics[width=\textwidth]{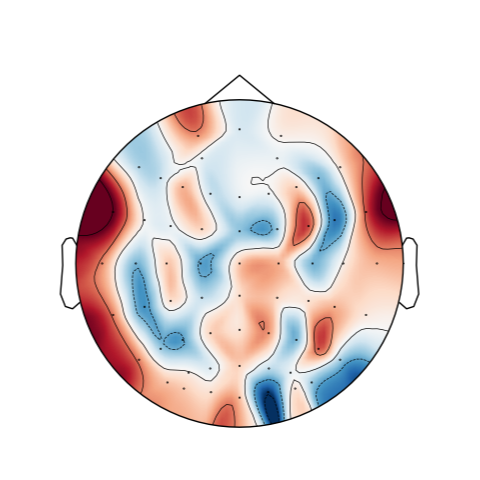}
    \caption{}
    \label{fig:image4}
\end{subfigure}
\hfill
\begin{subfigure}[b]{0.18\textwidth}
    \includegraphics[width=\textwidth]{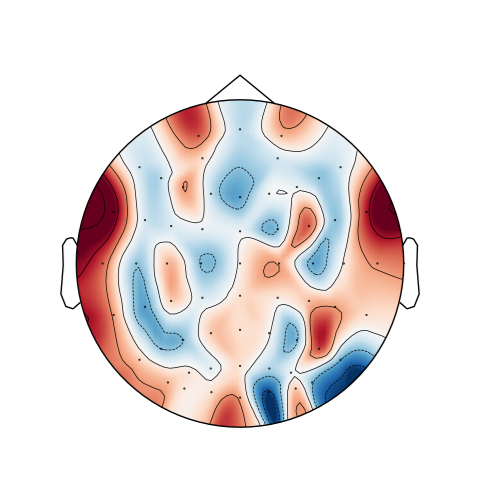}
    \caption{}
    \label{fig:image5}
\end{subfigure}

\captionsetup{belowskip=-10pt}
\caption{Attention visualization. We visualized the model's attention weights at different moments during the training process, which allows for an intuitive understanding of how the model's preference for EEG signals monitored by electrodes at various locations evolves over time (b, c, d, e). (a) presents the layout of the utilized 62-channel electrode placement.}
\label{fig:attn vlz}
\end{figure}

\subsection{Interpretability}
In order to provide a biologically plausible interpretation and inference for the research on EEG-based emotion recognition, as well as to substantiate the efficacy of the proposed model, we have conducted multiple visualizations of the attention outputs during the encoding process of EEG signals from test samples, projecting them back onto the electrode location map. This approach has allowed for a more intuitive validation of the brain regions associated with the task of emotion recognition, thereby enabling more insightful deductions.
\section{Experiment Result and Analysis}

\subsection{Datasets and Pre-processing}
Extensive experiments have been conducted on two public datasets, SEED and SEED-V. The SEED dataset includes EEG monitoring data and corresponding emotional labels (sad, happy, and neutral) from 15 subjects. Each subject completed 3 sessions, with each session comprising 15 trials, resulting in a total of 45 trials. Similarly, SEED-V was completed by 16 subjects, each participating in 45 trials, and includes EEG data along with corresponding labels for five emotions (disgust, fear, sad, happy, and neutral).

In the preprocessing of raw EEG data, a sequence involving a bandpass filter with cutoff frequencies of 0.1Hz and 70Hz, followed by a notch filter at 50Hz, was implemented. Subsequently, the sampling frequency was reduced to 200Hz from its original rate. For EEG segments corresponding to trials of varied lengths, a 4s non-overlapping Hanning window was utilized for segmentation in reverse order. The Short Time Fourier Transform (STFT) was then applied to calculate the DE feature across five frequency domains. Every four sliding window calculations were grouped together to extract EEG signals requisite for the pretrained model.

\subsection{Baseline Model and Settings}
We conducted subject-independent experiments on the SEED and SEED-V datasets using baseline models including DGCNN~\cite{song2018eeg}, RGNN~\cite{zhong2020eeg}, SOGNN~\cite{li2021cross} and BFE-Net~\cite{zhang2024subject}. Notably, baseline models that were trained using a single modality were explicitly annotated. The experimental framework utilized was PyTorch, with the GPU being NVIDIA A800 80GB PCIe. Furthermore, for each experiment, the data were randomly divided into training and testing sets in an 8:2 ratio. The model performance was evaluated based on the average accuracy and variance on the testing set.

\begin{table}
\centering

\captionsetup{belowskip=-10pt}
\caption{Subject-independent classification performance (Acc/Std\%) on SEED and SEED-V, where SWC represents whether to combine DE features in sliding window order.}
\label{tab:table1}
\begin{tabular}{ccccccccc} 
\toprule 
\multirow{3}{*}{Method} & \multicolumn{4}{c}{Modality} & \multicolumn{2}{c}{SEED} & \multicolumn{2}{c}{SEED-V} \\
\cline{2-9}
                        & \multicolumn{2}{c}{DE Feature} & \multirow{2}{*}{Raw EEG} & \multirow{2}{*}{Eye Movement} & \multirow{2}{*}{Acc}   & \multirow{2}{*}{Std}   & \multirow{2}{*}{Acc}   & \multirow{2}{*}{Std}   \\
\cline{2-3} 
& w/o SWC\,   & w SWC      &            &            &        &        &        &        \\
\midrule 
SVM         &\checkmark  &            &            &            &56.73\,     &16.29\,   &34.45       &13.67        \\
DGCNN       &\checkmark  &            &            &            &79.95\,     &09.02\,   &-           &-            \\
RGNN        &\checkmark  &            &            &            &85.30\,     &06.72\,   &66.28\,     &16.71\,      \\
SOGNN       &\checkmark  &            &            &            &86.81\,     &05.79\,   &74.53\,     &07.90\,      \\
BFE-Net     &\checkmark  &            &            &            &92.29\,     &04.65\,   &-           &-            \\
\midrule
SVM & &\checkmark            &  &    &70.23     &10.42     &56.49      & 12.51      \\
Mood reader & &\checkmark            & \checkmark &            &91.65     &05.42     &78.74      & 08.97      \\
Mood reader & &\checkmark            &            &\checkmark  &\textbf{93.12} &\textbf{04.75}&\textbf{84.36}&\textbf{05.23}       \\
\bottomrule 
\end{tabular}
\end{table}

\subsection{Results}

\subsubsection{Experiment Result}
Table~\ref{tab:table1} presents the experimental results of the baseline model and our method on the SEED and SEED-V datasets, with annotations regarding the categories of data utilized. The results demonstrate the consistently superior classification performance of Mood Reader across different datasets. Furthermore, they validate the efficacy of the sequentially combined DE features through a sliding window approach in the task of emotion recognition, as well as the correctness of the multimodal cross-scale information fusion strategy.

\subsubsection{Interpretation}
The results of attention visualization are summarized in Figure \ref{fig:attn vlz}. It can be observed that as the emotional recognition capability improves, the network's attention on EEG signals gradually shifts from a scattered global distribution to concentrated attention on specific regions. These brain regions include the frontal lobe area, areas of the left and right temporal lobes, and a small portion of the parietal lobe, which has been proven to be closely related to the generation and processing of emotions~\cite{dolcos2004interaction}.

Additionally, we noticed that there are also small areas within the occipital lobe, primarily responsible for visual information processing, that exhibit significant attention. Given that a substantial part of the stimuli in the experimental datasets SEED and SEED-V comes from visual stimuli in videos, we have reasonable grounds to propose the hypothesis that ``visually encoded information in the human brain is directly involved in emotion generation'' to a certain extent. This hypothesis also represents the holistic view that various parts of the brain participate in different functions and collectively process complex information~\cite{sporns2013structure}.

\subsection{Ablation Studies}
\begin{figure}[ht]
\centering
\includegraphics[width=0.8\linewidth]{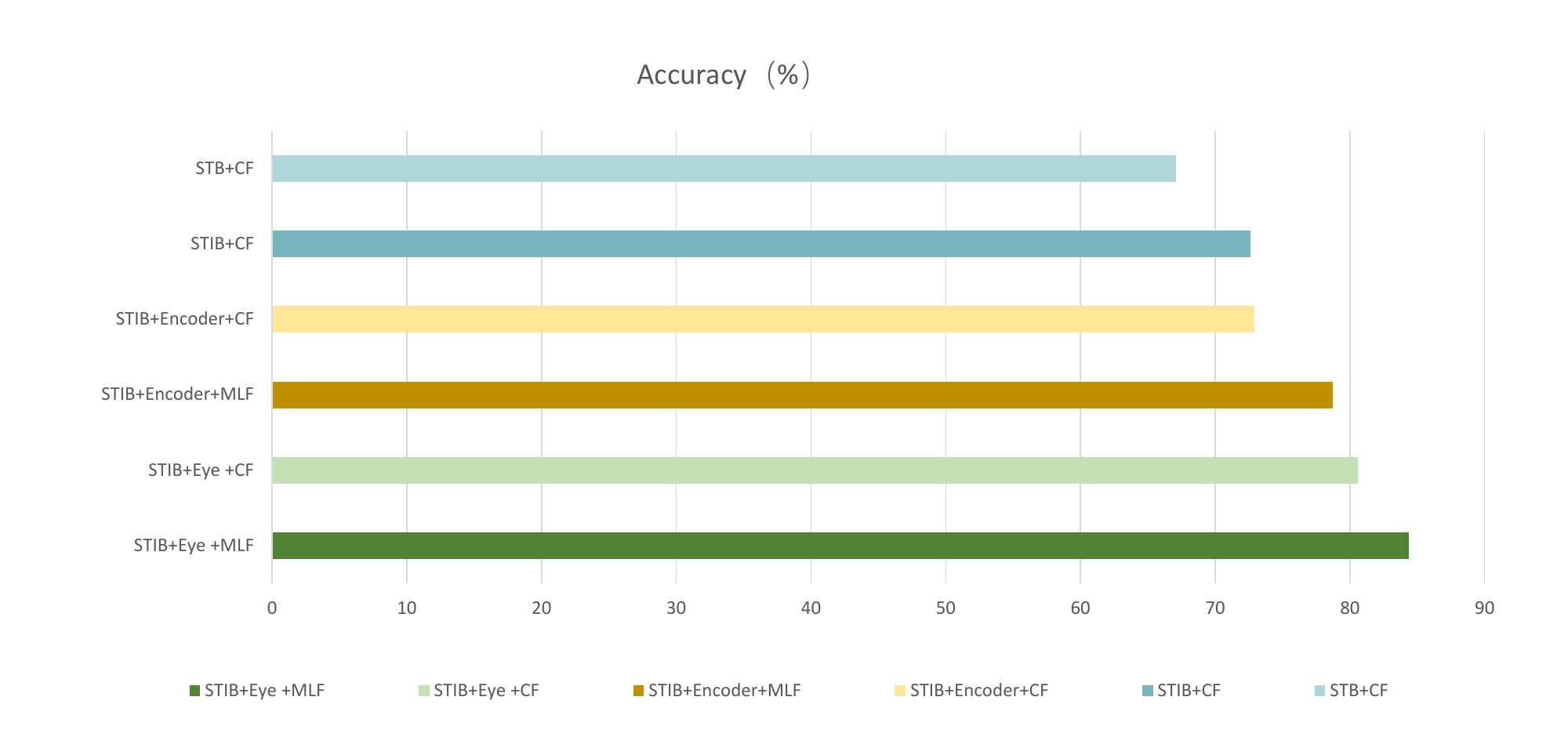}
\caption{The results of the ablation studies, conducted through progressive stacking of modules.}
\label{fig:ablation}
\end{figure}

To substantiate the effectiveness of the employed modules, we conducted ablation experiments on the SEED-V dataset using a stepwise stacking approach for the modules, with the specific experimental details as follows, and the results are depicted in Figure~\ref{fig:ablation}.
\begin{enumerate}
\item STB+CF: SWC DE with spatial-temporal block + concatenation fusion
\item STIB+CF: SWC DE with spatial-temporal interlinked block + concatenation fusion
\item STIB+Encoder+CF: SWC DE with spatial-temporal interlinked block + pre-trained encoder + concatenation fusion
\item STIB+Encoder+MLF: SWC DE with spatial-temporal interlinked block + pre-trained encoder + multi-level fusion
\item STIB+Eye+CF: SWC DE with spatial-temporal interlinked block + eye movement + concatenation fusion
\item STIB+Eye+MLF: SWC DE with spatial-temporal interlinked block + eye movement + multi-level fusion
\end{enumerate}
\vspace{-20pt}
\vspace{-20pt}

\section{Conclusion}
In this paper, we present the Mood Reader, a novel multimodal cross-scale fusion model for cross-subject emotion recognition based on EEG signals.Our model effectively integrates masked brain signal modeling for learning universal latent representations and an interlinked spatial-temporal attention mechanism to capture the complex dynamics of EEG signals. The multi-level fusion layer maximizes the advantages of features across different dimensions and modalities, leading to superior performance in cross-subject emotion recognition tasks. Furthermore, the model's interpretability, achieved through attention visualization, provides valuable insights into emotion-related brain areas, contributing to the understanding of neural processes underlying emotions. In conclusion, Mood Reader represents a significant step forward in cross-subject EEG-based emotion recognition, leveraging multimodal cross-scale fusion and advanced attention mechanisms.

\section*{Acknowledgement}

This work was supported in part by  the National Natural Science Foundations of China under Grant 62172403, the Distinguished Young Scholars Fund of Guangdong under Grant 2021B1515020019. M. Ng's research is supported in part by the HKRGC GRF 17201020 and 17300021, HKRGC CRF C7004-21GF, and Joint NSFC and RGC N-HKU769/21.

\bibliographystyle{splncs04}
\bibliography{ref}

\end{document}